# Cassini in situ observations of long duration magnetic reconnection in Saturn's magnetotail

C.S. Arridge[1,2,3], J.P. Eastwood[4], C.M. Jackman[5], G.-K. Poh[6], J.A. Slavin[6], M.F. Thomsen[7], N. André[8], X. Jia[6], A. Kidder[9], L. Lamy[10], A. Radioti[11], N. Sergis[12], M. Volwerk[13], A.P. Walsh[14], P. Zarka[10], A.J. Coates[15], M.K. Dougherty[4]

## Citation to the final published version

1. Department of Physics, Lancaster University, Bailrigg, Lancaster, LA1 4YB, United Kingdom.

2. Mullard Space Science Laboratory, University College London, Holmbury St. Mary, Dorking, Surrey, RH5 6NT, United Kingdom.

3. The Centre for Planetary Science at UCL/Birkbeck, Gower Street, London, WC1E 6BT, United Kingdom.

4. Department of Physics, Imperial College, South Kensington, London, SW7 2BW, United Kingdom.

5. School of Physics and Astronomy, University of Southampton, Southampton, SO17 1BJ, United Kingdom.

6. Department of Atmospheric, Oceanic and Space Sciences, University of Michigan, 2455 Hayward St., Ann Arbor, Michigan 48109-2143, USA

7. Planetary Science Institute, 1700 East Fort Lowell, Suite 106, Tucson, Arizona 85719-2395, USA.

8. CNRS, Institut de Recherche en Astrophysiqe et Planétologie, 9 avenue du colonel Roche, BP 44346, 31028 Toulouse Cedex 4, France.

9. Department of Earth and Space Sciences, University of Washington, Box 351310, Seattle, Washington 98195, USA.

10. LESIA-Observatoire de Paris, CNRS, UPMC Univ. Paris 6, Univ. Paris-Diderot, 92195, Meudon, France.

11. Laboratoire de Physique, Atmosphérique et Planétaire, Institut d'Astrophysique et de Géophysique, Université de Liège, Liège, Belgium.

12. Office for Space Research, Academy of Athens, 4, Soranou Efesiou str., 11527, Papagos, Athens, Greece.

13. Austrian Academy of Sciences, Space Research Institute, Schmiedlstraße 6, 8042 Graz, Austria.

14. Science and Robotic Exploration Directorate, European Space Agency, ESAC, Villanueva de la Cañada, 28692 Madrid, Spain.

15. Mullard Space Science Laboratory, Department of Space and Climate Physics, University College London, Holmbury St. Mary, Dorking, Surrey, RH5 6NT, United Kingdom.

## First paragraph

Magnetic reconnection is a fundamental process in solar system and astrophysical plasmas through which stored magnetic energy associated with current sheets is converted into thermal, kinetic and wave energy. Magnetic reconnection also ought to be a key process involved in shedding internally-produced plasma from the giant magnetospheres at Jupiter and Saturn through topological reconfiguration of the magnetic field. Here we report the first encounter of the Cassini spacecraft with an ion diffusion region in Saturn's magnetotail, and additional signatures of magnetic reconnection over 19 hours. This directly reveals reconnection in fact can act for prolonged intervals (in excess of the planetary rotation period) and is a significant pathway for internal plasma loss at Saturn, countering the view of reconnection as a transient method of internal plasma loss at Saturn. These results have application to understanding the rapidly rotating magnetosphere at Jupiter and also to other rapidly rotating astrophysical magnetospheres.

## Main

Since the discovery of $H_2O$ plumes from the icy moon Enceladus it has become clear that the dominant source of plasma in Saturn's magnetosphere is the ionisation of neutral molecules deep within the magnetosphere producing a plasma composed of $H_2O^+$, $H_3O^+$, $OH^+$, collectively referred to as the water group, $W^+$ (1-3). Some of this plasma is lost from the system by charge-exchange, the remaining plasma is transported radially outward. The radial transport is driven by the centrifugal interchange instability, which is analogous to the Rayleigh-Taylor instability with gravity replaced by the centrifugal force associated with the rapid rotation of the magnetosphere[4]. Magnetic reconnection is a process involving topological rearrangement of the magnetic field which results in the closure of magnetic flux opened at the dayside magnetopause, and also results in mass loss from the magnetosphere. In a time-averaged state the outward plasma transport rate should match the plasma loss rate through magnetotail reconnection, and the dayside reconnection rate should match that in the magnetotail. Observations of magnetotail reconnection thus provide a method to test the loss process for this internally-produced plasma, as well as the closure of magnetic flux opened at the magnetopause.

Data from the Cassini spacecraft has only provided indirect evidence for magnetotail reconnection[5,6,7,8] but the actual region where magnetic fields are merging, known as the diffusion region, has not been detected at Saturn, or Jupiter. The diffusion region has a two-scale structure with the larger ion diffusion region surrounding the smaller electron diffusion region. The ion diffusion region has been detected in observations in Earth's magnetotail[9,10,11], Earth's magnetosheath[12], the solar wind, and at Mars[13,14]. The plasma loss rates inferred from these previous observations of magnetic reconnection at Saturn and Jupiter are an order of magnitude too small when compared to the known plasma production rates[8,15,16]. Here we report the first observations of an ion diffusion region in Saturn's magnetotail. These direct observations show that reconnection can occur over prolonged intervals, almost an order of magnitude longer than the longest previously reported[17].

Figure 1 shows magnetic field and electron data for a six hour interval on 08 October 2006 when Cassini was located in the post-midnight sector of Saturn's magnetosphere around 0130 Saturn Local Time, about 8º north of Saturn's equatorial plane, and at a radial distance of 29 $R_S$, where 1 $R_S$=60268 km. As illustrated in Figure 2, the magnetic field in the tail is generally in a swept-back configuration as the result of outward plasma transport and angular momentum conservation. This effect is removed by rotating the data into a new right-handed coordinate system where the background magnetic field is in the X direction, and the Y direction is perpendicular to the plane of the swept-back magnetic field (details of the transformation are given in the Supplementary Material). At the beginning of the interval, Cassini was located above the magnetotail current sheet ($B_x>0$), crossing below ($B_x<0$) the centre of the current sheet between 03:30 UT – 03:40 UT. $B_z$ is ordinarily expected to be negative. At 03:55 UT $B_z$ reverses sign, which in fact corresponds to Cassini crossing the X-line from the tailward to the planetward side as shown in Figure 2. The quantity $|B_z|/\max(|B_x|)$ is an estimate of the reconnection rate and was found to be 0.13±0.10 with a peak of 0.66 – hence consistent with fast reconnection[13].

On the tailward side of the X-line a very energetic (~10 keV/q) ion population is observed flowing tailward, and slightly duskward. This population is not a field-aligned ion beam and is convective (i.e. a has significant perpendicular velocity component). These ions are moving with speeds of 1200 km s$^{-1}$, a substantial fraction of the lobe Alfvén speed of ~4000 km s$^{-1}$ (19) and much larger than corotation speed (~150 km s$^{-1}$), and are identified as the tailward jet from the X-line. On the planetward side of the X-line the field-of-view of CAPS does not cover the region where we would expect to see planetward ion beams. Later that day as Cassini leaves the diffusion region the plasma flow returns to near corotation, but with a tailward and northward component. Detailed analysis of these ion flow directions are presented in the Supplementary Material.

Around the magnetic reconnection site ideal magnetohydrodynamics breaks down and charged particles become demagnetised from the magnetic field. Because of factor of ~1800 in the mass difference between electrons and ions, the ions demagnetise over a large spatial region than electrons resulting in differential motion between ions and electrons. The resulting current system is known as the Hall current system and produces a characteristic quadrupolar magnetic field structure in the out-of-plane magnetic field, $B_y$ (Figure 2). The red (blue) regions of $B_x$ and $B_y$ in Figure 1 indicate where the $B_y$ component is expected to have a positive (negative) sign associated with this current system and this colour-coding is consistent with the Hall field. As expected, the strength of the Hall field perturbation peaks between the centre of the current sheet and the lobe. All four quadrants of the Hall field were measured by Cassini. As calculated in the Supplementary Material, the strength of the Hall field can be estimated by the quantity $|B_y|/\max(|B_x|)$ and the mean value of 0.18±0.15 is somewhat smaller than that observed in other environments although the peak of 0.83 is more consistent with the typical strength, ~0.5, of the Hall field[9,14].

As shown in the Supplementary Material, further evidence for the detection of the ion diffusion region is found in the form of secondary tearing islands (small loop-like magnetic field structures) at 02:20-03:00 UT and 03:28-03:40 UT, cool electrons flowing in response to the Hall current system, and hot electrons flowing out of the X-line. Taken together, the unambiguous conclusion is that Cassini encountered a tailward retreating X-line and ion diffusion region in Saturn's magnetotail as sketched in Figure 2. As shown in the Methods, the ion diffusion region is $>\sim 0.1$ $R_S$ (~6000 km) in size; the lower plasma density in the saturnian system means that the ion diffusion region is an order of magnitude larger than at Earth. Cassini spends over 150 minutes near the reconnection site, which although is longer than ~10 minutes at Earth, is not unexpected given the differing size of the diffusion region itself.

In the Methods we have estimated that 0.34 GWb $R_S^{-1}$ of magnetic flux is closed during this diffusion region encounter, where the dimensions include per unit length because the length of the X-line is unknown. Previous work assumes an upper limit of 90 $R_S$ (8) and hence 31 GWb of magnetic flux is closed. This is more than an order of magnitude greater than the largest estimates from indirect reconnection observations alone[8]. Our observations are entirely consistent with rates of flux closure inferred from auroral observations[22,23]. Similarly, we estimate the mass lost during this diffusion region encounter to be $3\times10^5$ kg $R_S^{-1}$ and hence $3\times10^7$ kg, three orders of magnitude larger than previous estimates[17]. Events of this magnitude every ~4-40 days are required to match a time-averaged mass loading rate of 100 kg/s, rather than every 7 minutes from previous estimates based on in direct observations[16]. Hence, these results demonstrate that magnetotail reconnection can close sufficient amounts of magnetic flux and act as a very significant mass loss mechanism.

Additional indirect signatures of magnetic reconnection are also observed two hours after the X-line retreats tailward. Figure 3 shows five hours of electron fluxes and magnetometer data revealing a series of reconnection signatures in a spherical polar (Kronocentric radial-theta-phi, KRTP) coordinate system. Bipolar perturbations in the $B_\theta$ component indicate the passage of a plasmoid, a loop-like magnetic flux structure[8]. At 0605 UT a tailward moving plasmoid passes the spacecraft, sourced from an X-line planetward of the spacecraft. At 0705 and 0810 UT a sharp increase in $B_\theta$ to large positive values is indicative of a dipolarisation front, the compression of magnetic field lines around plasma moving rapidly towards the planet as the result of magnetic reconnection downtail from the spacecraft[24]. These two dipolarisation fronts indicate the presence of an X-line tailward of the spacecraft. Following the passage of the fronts the spacecraft is immersed in hot plasma, similar to that seen in Earth's magnetotail[25], and is a signature of the energy conversion in the reconnection process. After the final dipolarisation front passes Cassini, the spacecraft is located in a region of fluctuating magnetic fields similar to a chain of magnetic islands and is surrounded by energetic ~10 keV electrons[26] which also display evidence of becoming more energetic with time. Ion flows with a planetward component are found throughout this hot plasma region with speeds in excess of ~1000 km s$^{-1}$. Towards the end of the interval, between 15:00 and 17:25 UT, planetward flowing ions and electrons are found in a layer between the plasma sheet and the lobes, which are consistent with outflows from a more distant X-line[27]. The detailed particle analysis is presented in the Supplementary Material.

These data are evidence for ongoing but time variable magnetic reconnection in the magnetotail at this local time over a period of 19 hours, covering almost two rotations of Saturn. Simulations of upstream solar wind conditions presented in the Supplementary Information show that the magnetosphere was strongly compressed just before the entry into the X-line, suggesting triggering of tail reconnection by a solar wind pressure pulse. As shown in the Supplementary Material, a weaker pressure pulse arrives on 09 October at 1400 UT when Cassini was located in the inner magnetosphere. Wave signatures suggest that this triggered further reconnection. These observations stand in contrast to the much less frequent plasmoid observations that have previously been used to infer rates of magnetic reconnection in Saturn's magnetotail. At this point it is not possible to determine whether this is a consequence of the magnitude of the solar wind pressure increase, or if this is simply a common event but rarely observed due to the orbit of Cassini and the spatial distribution/spatial size of diffusion regions. These results show that prolonged magnetotail reconnection can close sufficient magnetic flux and shed sufficient mass to explain the time-averaged driving of Saturn's magnetosphere.

## Methods

### Magnetic flux closure and plasma evacuation

Plasmoids have been estimated to close between 0.26 and 2.2 GWb of magnetic flux, by integrating the product of the $B_\theta$ component of the magnetic field and the tailward flow speed, with an assumption that the reconnection extended 90 $R_S$ across the whole tail[8]. Applying the same argument to the data in the ion diffusion region in figure 1 and a flow speed of 1200 km s$^{-1}$, (based on the ion measurements), the reconnected flux is 0.34 GWb $R_S^{-1}$ or 31 GWb assuming the width of 90 $R_S$ This is more than an order of magnitude greater than the largest estimates based on plasmoid observations alone[8]. From changes in the size of Saturn's main auroral oval, changes in open tail flux are typically 5 GWb over a 10-60 hour period[21] but, occasionally, can be much higher (3.5 GWb/hour) (22). These are consistent with our observations.

By scaling the rate of flux closure by the mass per unit flux ~10$^{-3}$ kg/Wb (20), we estimate that this releases 3×10$^5$ kg $R_S^{-1}$ or 3×10$^7$ kg. Estimates of the mass lost per plasmoid can be made by combining typical tail plasma densities with an estimate for the plasmoid volume, to give 62×10$^3$ kg per plasmoid. Hence, ~200 plasmoids per day (one every ~7 minutes) are required to remove the plasma transported outwards from the inner magnetosphere[16]. Hence, our mass loss rate is three orders of magnitude larger. Events of this magnitude every ~4-40 days are required to match a time-averaged mass loading rate of 100 kg/s.

**Size of the ion diffusion region**

In two-fluid reconnection the size of the ion diffusion region is a multiple of the ion inertial length, $c/\omega_i$, where c is the speed of light in a vacuum and $\omega_i$ is the ion plasma frequency given by $(nZ^2e^2/\varepsilon_0 m_i)^{1/2}$, where n is the ion number density, Z is the ion charge state, e is the fundamental charge, $\varepsilon_0$ is the permittivity of free space, and $m_i$ is the ion mass. Using measurements of tail plasma at 30 $R_S$ with a plasma number density of $4 - 8 \times 10^4$ $m^{-3}$ and composition of $n_{W+}/n_{H+}$~2 (20), the mean ion mass is $1.95 \times 10^{-26}$ kg (~$11.7 m_p$) and the ion inertial length is 3000 – 4000 km (0.05 – 0.06 $R_S$).

# Acknowledgements

CSA was funded in this work by a Royal Society Research Fellowship and an STFC Postdoctoral Fellowship. CMJ was funded by an STFC Ernest Rutherford Fellowship. MFT was supported by the NASA Cassini program through JPL contract 1243218 with Southwest Research Institute and is grateful to Los Alamos National Laboratory for support provided her as a guest scientist. CSA/CMJ/JAS/NA/XJ/AK/AR/MV/APW acknowledge the support of the International Space Science Institute where part of this work was carried out. Cassini operations are supported by NASA (managed by the Jet Propulsion Laboratory) and by ESA. The data reported in this paper are available from the NASA Planetary Data System. SKR data were accessed through the Cassini/RPWS/HFR data server http://www.lesia.obspm.fr/kronos developed at the Observatory of Paris/LESIA with support from CNRS and CNES.

# Figure captions

Figure 1: Interval encompassing an ion diffusion region in Saturn's magnetotail as seen by the Cassini spacecraft. Panel (a) electron omnidirectional flux time-energy spectrogram in units of differential energy flux (eV m$^{-2}$ sr$^{-1}$ s$^{-1}$ eV$^{-1}$); (b-d) three components of the magnetic field in the X-line coordinate system, parts of the $B_x$ and $B_y$ traces in red (blue) show where the $B_y$ component is expected to be positive (negative); (e) the field magnitude.

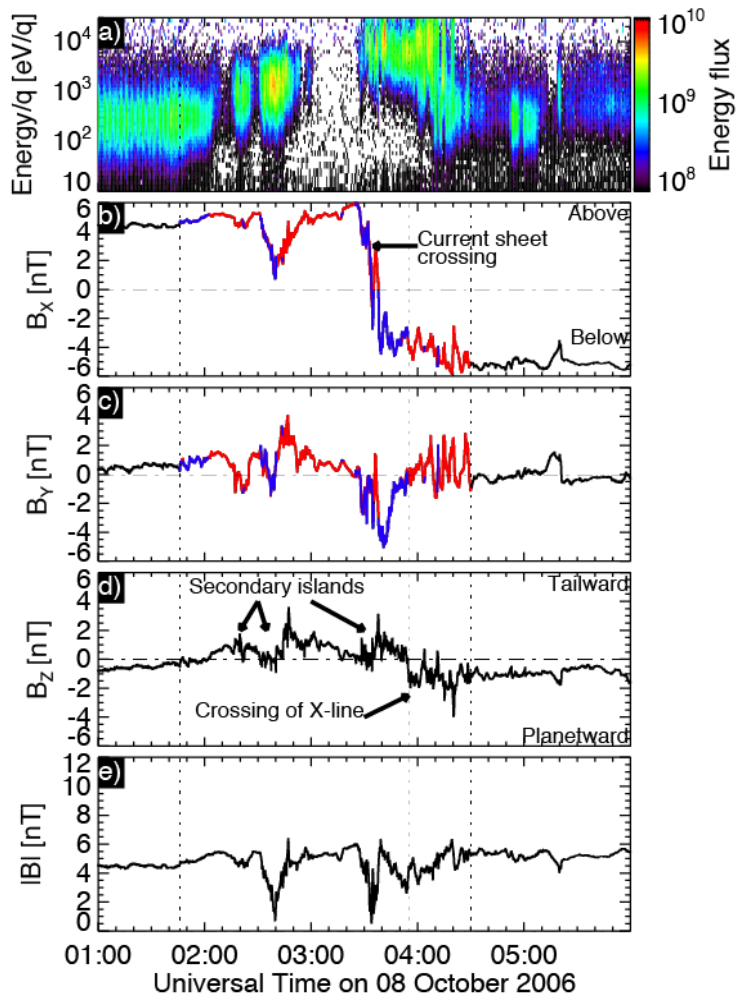

Figure 2: Geometry of the X-line coordinate system and schematic of Cassini's motion relative to the X-line. The red vectors show the original spherical polar coordinate system from the magnetometer data and the green vectors show the new X-line coordinate system which takes into account the swept-back configuration of the magnetic field. The blue curve in the top two panels shows the orbit of Cassini around Saturn and in the bottom view we show the inferred motion of Cassini relative to the magnetic reconnection X-line. The pink and blue regions are the ion and electron diffusion regions[9].

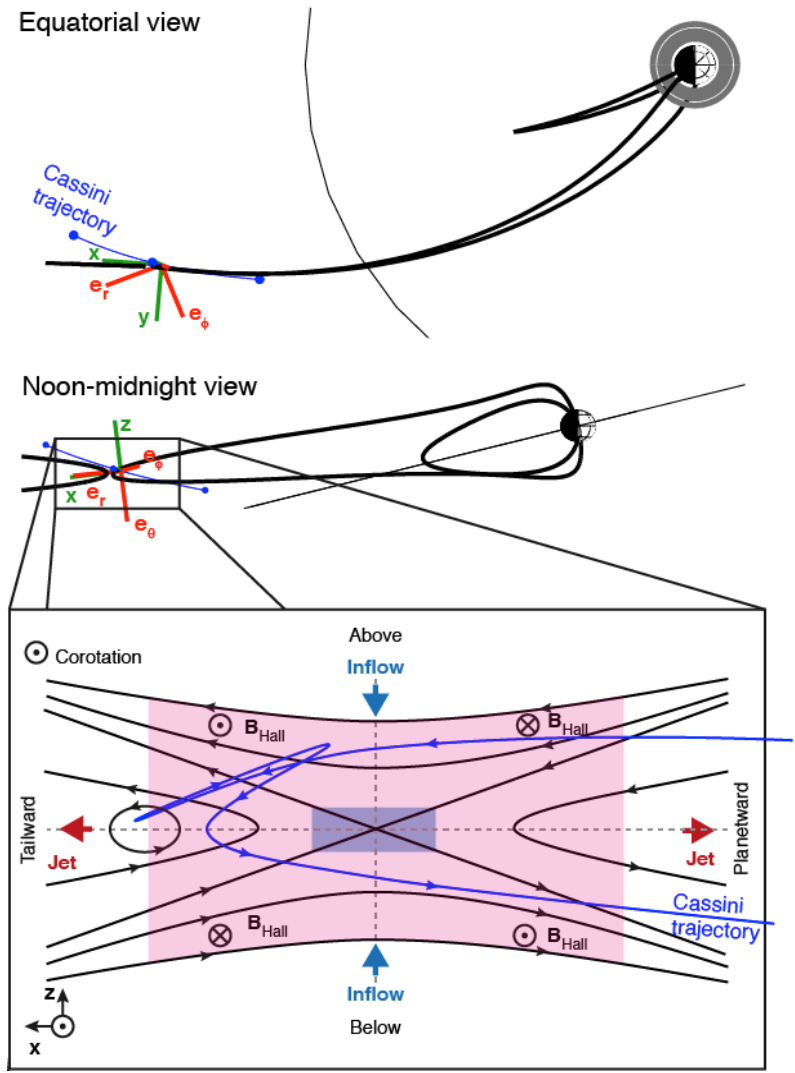

Figure 3: Dipolarisation fronts (DF), plasmoid (P), and the restart of reconnection. Panel (a) electron omnidirectional flux time-energy spectrogram in units of differential energy flux (eV m$^{-2}$ sr$^{-1}$ s$^{-1}$ eV$^{-1}$); (b-d) three components of the magnetic field in spherical polar coordinates. The grey region indicates periods where the spacecraft is immersed in the plasma sheet.

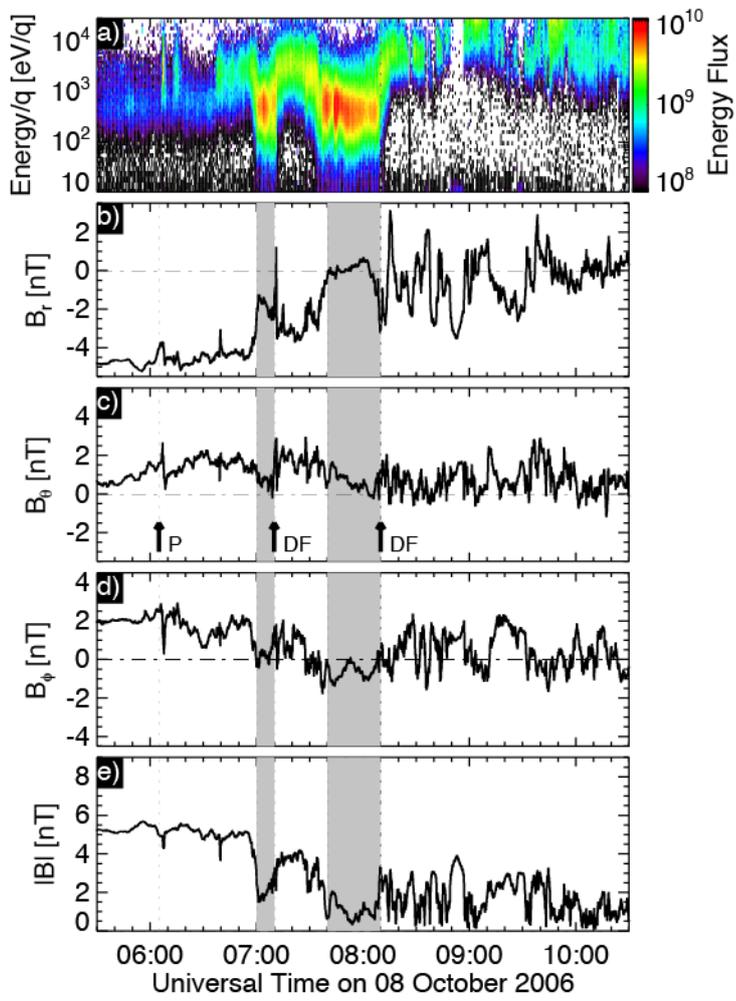

# Supplementary Material for:

# Cassini in situ observations of long duration magnetic reconnection in Saturn's magnetotail

C.S. Arridge[1,2,3], J.P. Eastwood[4], C.M. Jackman[5], G.-K. Poh[6], J.A. Slavin[6], M.F. Thomsen[7], N. André[8], X. Jia[6], A. Kidder[9], L. Lamy[10], A. Radioti[11], N. Sergis[12], M. Volwerk[13], A.P. Walsh[14], P. Zarka[10], A.J. Coates[15], M.K. Dougherty[4]

1. Department of Physics, Lancaster University, Bailrigg, Lancaster, LA1 4YB, United Kingdom.

2. Mullard Space Science Laboratory, University College London, Holmbury St. Mary, Dorking, Surrey, RH5 6NT, United Kingdom.

3. The Centre for Planetary Science at UCL/Birkbeck, Gower Street, London, WC1E 6BT, United Kingdom.

4. Department of Physics, Imperial College, South Kensington, London, SW7 2BW, United Kingdom.

5. School of Physics and Astronomy, University of Southampton, Southampton, SO17 1BJ, United Kingdom.

6. Department of Atmospheric, Oceanic and Space Sciences, University of Michigan, 2455 Hayward St., Ann Arbor, Michigan 48109-2143, USA

7. Planetary Science Institute, 1700 East Fort Lowell, Suite 106, Tucson, Arizona 85719-2395, USA.

8. CNRS, Institut de Recherche en Astrophysiqe et Planétologie, 9 avenue du colonel Roche, BP 44346, 31028 Toulouse Cedex 4, France.

9. Department of Earth and Space Sciences, University of Washington, Box 351310, Seattle, Washington 98195, USA.

10. LESIA-Observatoire de Paris, CNRS, UPMC Univ. Paris 6, Univ. Paris-Diderot, 92195, Meudon, France.

11. Laboratoire de Physique, Atmosphérique et Planétaire, Institut d'Astrophysique et de Géophysique, Université de Liège, Liège, Belgium.

12. Office for Space Research, Academy of Athens, 4, Soranou Efesiou str., 11527, Papagos, Athens, Greece.

13. Austrian Academy of Sciences, Space Research Institute, Schmiedlstraße 6, 8042 Graz, Austria.

14. Science and Robotic Exploration Directorate, European Space Agency, ESAC, Villanueva de la Cañada, 28692 Madrid, Spain.

15. Mullard Space Science Laboratory, Department of Space and Climate Physics, University College London, Holmbury St. Mary, Dorking, Surrey, RH5 6NT, United Kingdom.

# Trajectory

Figure S1 shows the trajectory of Cassini in Kronocentric Solar Magnetospheric (KSM) coordinates, where the X axis points from Saturn to the Sun, the X-Z plane contains the spin axis of Saturn, and Y points towards dusk. After periapsis with Saturn on 25 September 2006, Cassini started its orbit (Revolution) 30 and moved out into the magnetotail via dusk reaching an apoapsis of 36.6 $R_S$ on 03 October 2006 at 1904 UT at a local time of 00h18m and latitude of 20.9º with the spacecraft moving towards the equator. At the start of the reconnection event at 0146 UT on 08 October 2006 (day of year 281) the spacecraft was at 29.0 $R_S$, a latitude of 9.25º and local time 01h27m. The KSM coordinates at the start of the reconnection event was (-26.8, -10.6, -2.63) $R_S$. From figure S1 we can see that the spacecraft was located slightly north of the warped magnetospheric current sheet as can also be seen in the observations (figure 1).

# Instrumentation

Data in this study comes from the magnetometer, plasma spectrometer (CAPS), Radio and Plasma Wave Science (RPWS), and Magnetospheric Imaging Instrument (MIMI) instruments on the Cassini spacecraft. Upstream solar wind conditions are obtained from the ENLIL model[1] and are discussed in more detail in the next section.

Magnetometer data are taken from the fluxgate magnetometer instrument at a cadence of 1s in a spherical polar coordinate system centred on the spacecraft (Kronographic Radial-Theta-Phi, KRTP) which is based on the kronographic position of the spacecraft, where the radial vector, $e_r$, is oriented from the planet to the spacecraft, the polar vector, $e_\theta$, points in the direction of increasing co-latitude, and the azimuthal vector $e_\varphi$ completes the right-handed set and is oriented in a prograde direction around Saturn.

Plasma data are taken from the CAPS electron spectrometer (ELS) and ion mass spectrometer (IMS) which are electrostatic analysers but where IMS also has a time-of-flight (TOF) section to determine the energy-resolved mass per charge ratio of the incoming ions with a mass/charge resolution of 12.5%. ELS detects electrons between 0.6 and 28750 eV/e in 63 energy bins with a resolution of ΔE/E of 16.7%. The instantaneous field-of-view (FOV) is split into eight 20°×5.2° anodes providing a total 160°×5.2° instantaneous FOV. ELS sweeps this FOV every 2 s but these samples can be averaged on board to lower time and energy resolution. IMS detects positive ions between 1 and 50280 eV/q in 63 energy bins with a resolution of ΔE/E of 16.7% and a cadence of 4 s. Similar to ELS, the instantaneous FOV is split into eight anodes each with an FOV of 20°×8.3° providing a total instantaneous FOV of 160°×5.3°. The FOV of ELS and IMS are approximately boresighted. To improve the FOV the whole CAPS instrument is mounted on a rotating platform which sweeps the sky by around 1°/s, extending the FOV to ~$2\pi$ sr with a period of ~3 minutes. The spacecraft was also rolling for part of the interval reported in this paper which improves the total field-of-view to almost 4π sr but complicates the analysis as described in the appropriate sections below.

Radio data is provided by the RPWS instrument which includes three nearly orthogonal electric field antennae to detect AC electric fields between 1 Hz and 16 MHz and are particularly processed in this paper to analyse kilometric radio emissions[2].

## Solar wind simulations and Cassini remote sensing observations

Since there is no upstream monitor at Saturn models must be used to understand the upstream solar wind and interplanetary magnetic field (IMF) conditions while the spacecraft is inside the magnetosphere, as it was during this event. The MSWiM model is a 1.5-d MHD propagation of solar wind conditions measured at 1 AU but is only usable near apparent opposition which occurred on 25 February 2006. During the October 2006 time period Saturn is far from apparent opposition and so this model is not reliable. ENLIL is a 3D MHD simulation of the heliosphere[1] which is available at the Community Coordinated Modeling Center (CCMC) at NASA Goddard Space Flight Center. This model is not hampered by the same opposition viewing effects as MSWiM. The model inner boundary condition is provided by coronal models, driven by observed magnetograms, and is placed at 21.5 or 30 solar radii depending on the coronal model. Although limited validation studies of ENLIL have been performed for the outer heliosphere near Saturn, uncertainties on the arrival times for stream interaction regions can be up to four days at 5 AU, from a comparison of ENLIL results with Ulysses data[3]. In this work, version 2.7 of ENLIL was run with an inner boundary condition provided by the Wang-Sheely-Arge model for Carrington rotation 2048 and provided solar wind simulation results at Saturn's position from 21 September to 24 October 2006. In order to properly compare the in situ Cassini data with the ENLIL results we use Cassini observations of auroral radio emissions (Saturn Kilometric Radiation, SKR), known to brighten in response to solar wind compress[4,5]. These observations are used to identify a time shift that can be applied to the ENLIL results.

Figure S2 contains a summary of Cassini radio and plasma wave observations and ENLIL solar wind simulations for the period covering the event. The unshifted ENLIL data is shown in blue and the shifted data (discussed below) is in black. The interval encompasses a corotating interaction region (CIR) where the pressure and magnetic field strength increase. Four crossings of the heliospheric current sheet (HCS) are identified from reversals in the $B_T$ component of the magnetic field in Radial-Tangential-Normal (RTN) coordinates. Such crossings are typically embedded within CIR compressions at Saturn. The presence of a forward shock (FS) from an increase in the solar wind speed and a coincident increase in the dynamic pressure is also identified.

Turning to the Cassini Radio and Plasma Wave (RPWS) data in figures S2a and S2b: prior to the event on 08 October the flux density displays periodic increases in flux as commonly found in Saturn's magnetosphere[2] and has a right-hand circular polarisation consistent with extraordinary mode emission from the northern hemisphere, as expected from Cassini's northern latitude (figure S2c). These periodic emissions are found to occur at the expected phase for northern SKR emissions, labelled N at the top of figure S2a (6).

The white arrows in figure S2a identify example enhancements in SKR flux density with associated low frequency extensions (LFE) and a right-hand circular polarisation (northern hemisphere emission). The physical significance of these LFEs has been linked to increased precipitation of particles into the auroral zone and growth/movement of the radio source to higher altitudes (and hence lower frequencies since the emission frequency is inversely proportional to magnetic field strength). For example, at 0800 UT on 29 September, 1200 UT on 05 October and 2000 UT on 06 October, and occur at or near the expected phase for northern hemisphere emissions and are characteristic of internally-triggered SKR enhancements controlled by magnetospheric rotational modulation[7].

Following these LFEs there are two long-lasting enhancements in SKR power on 08 October for 15 hours and 11 October for 24 hours, more characteristic of an external solar wind control[4]. During these periods SKR is a very strong emission that lasts for more than one Saturn rotation, and does not have any correlation with northern or southern SKR phase[6]. The low frequency range (<10 kHz) displays intense SKR. The disappearance of SKR emissions around 2300 UT on 12 October is due to the spacecraft reaching Saturn periapsis (e.g., Figure S3) where SKR is not visible. The detached nature of the low frequency SKR emissions may be produced by to a spatially separated (in longitude or latitude) source region with different regions producing the high- and low-frequency emissions. A more likely interpretation is that the gap is due to refractive effects, supported by the abrupt change in SKR polarisation near 70-80 kHz which is quite readily understood in terms of refraction. Unfortunately it was not possible to apply direction-finding techniques to this data interval to be more certain on the origin of the detached low-frequency emissions.

The first event originates from the northern hemisphere (right-hand circular polarisation) and the second from the southern hemisphere (left-hand circular polarisation). If these were the same event, but viewed from the northern, then the southern hemisphere, we might expect to see a change in polarisation at the equator. However, the northern hemisphere emission fades well before the spacecraft crosses the equator, and at a point where the latitude and local time are varying slowly. The near-equatorial spacecraft location during these two events discards strong visibility effects[2]. Hence, this is evidence for two periods of long-lasting SKR enhancement that are driven separately by external large-scale compressions of the magnetosphere. Therefore we associate these two periods of strong SKR emissions with external compressions of the magnetosphere. We shifted the ENLIL time-series by 5.3 days such that the first major SKR enhancement begins at the arrival of the first large pressure pulse in the ENLIL time-series. This was done by matching the rise in dynamic pressure with the rise in intensity of SKR emissions. Given the ~10 hour lag between the arrival of a solar wind dynamic pressure front and the increase in SKR emissions[5] we assign an uncertainty of 0.5 days to this estimate (4.8-5.3 days). In doing this, the second strong enhancement in SKR flux density matches the second pressure pulse in the ENLIL results thus providing supporting evidence that these enhancements in SKR are associated with externally-driven magnetospheric compressions. We also note that the increase in solar wind dynamic pressure occurs at the forward shock (FS) and occurs approximately at the same time as the onset of the periodic LFEs and the onset of this activity might represent the arrival of the CIR at Saturn.

Finally, the low frequency SKR emissions are accompanied by rising periodic narrow band emissions, mainly with opposite polarisation. These appear at frequencies around 5 kHz, so-called Saturnian Myriametric Radiation or n-SMR (*8*) and around 20 kHz, identified as narrowband SKR or n-SKR[2]. n-SMR are similar to continuum emissions from Earth's plasmapause, and n-KOM emissions from the Io torus, which are known to be generated at density gradients[8]. These might be attributed to dynamics internal to the plasma disc but in this case there is evidence that they are triggered by increases in the solar wind dynamic pressure. Although the spacecraft is moving latitudinally, there is no correlation of the morphology of the emissions with the location of the spacecraft, and the emissions appear after the major magnetospheric compressions (figure S2g). Activity in n-SKR and n-SMR continues however until 17 October, which is a much longer period than the 4 – 5 days previously reported[8] and may reflect the strength of the external compression, or that the initial external trigger has resulted in a "cascade" of internally-driven responses.

In summary, shifting the ENLIL time series by 4.8-5.3 days (to form the shifted time series in figure S2) we arrive at the following sequence of upstream events. Between 0000 UT and 1200 UT on 06 October a forward shock impacted Saturn and over the course of ~12 hours the magnetosphere was slowly compressed from a subsolar magnetopause position of 25 $R_S$ to 17 $R_S$ representing a moderate compression due to the enhanced compressibility of Saturn's magnetosphere compared to Earth[9]. A pressure pulse with a peak dynamic pressure of 0.23 nPa arrives between 1200 UT on 07 October and 0000 UT on 08 October compressing the magnetosphere over the next ~6 hours such that the magnetopause subsolar distance decreases to 14±2 $R_S$, representing an extreme and relatively rare compression. The pressure pulse begins to fade around 16 hours after it arrived falling back to a magnetopause subsolar distance of ~20 $R_S$ by the end of 08 October. Between the middle of the day on 09 October and early on 10 October a smaller pressure pulse arrives producing a magnetopause standoff distance of 16±2 $R_S$.

## Rotation of the magnetic field data to remove the effect of sweepback

The magnetic field at Saturn is swept-back into a lagging configuration over most local times produced by a combination of magnetopause currents and outward transport of internally produced plasma[10], although the latter is thought to dominate the observed sweep-back. The effect of this sweep-back is to introduce an azimuthal component to the magnetic field (in spherical polar coordinates) which reverses in sense about the centre of the current sheet such that the azimuthal and radial components of the field have an anti-phase relationship. Typically, $B_r>0$ and $B_\varphi<0$ above the current sheet, and $B_r>0$ and $B_\varphi<0$ below the current sheet. In collisionless reconnection, separation of ions and electrons occurs as the ions demagnetise in the ion diffusion region but where the electrons remain frozen to the field and continue to inflow towards the X-line where they eventually demagnetise at the electron scale. This separation of ions and electrons produces a current system known as the Hall current and associated field (the Hall field)[11]. The Hall field has a quadrupolar structure with out-of-plane components.

Figure S4 illustrates the relationship between the Hall field and the azimuthal field associated with sweep-back and highlights the fact that the presence of the Hall field may be masked by the swept-back configuration of the field. For example, on the planetward side of the X-line the Hall field has a negative out-of-plane component above the current sheet but the swept-back configuration also produces a negative out-of-plane component.

Hence, in the KRTP coordinate system it is hard to detect the presence of the Hall field. To clearly identify the Hall field we rotate the magnetic field data into an X-line coordinate system using the sweep-back angle of the field, defined as $\alpha=\tan^{-1}(B_\varphi/B_r)$:

$$\begin{pmatrix} B_x \\ B_y \\ B_z \end{pmatrix} = \begin{pmatrix} \cos\alpha & 0 & \sin\alpha \\ -\sin\alpha & 0 & \cos\alpha \\ 0 & -1 & 0 \end{pmatrix} \begin{pmatrix} B_r \\ B_\theta \\ B_\varphi \end{pmatrix} \qquad 1$$

This produces an X-line coordinate system where X points approximately tailward, Z points approximately northward, and Y completes the right-handed set pointing approximately dawnward. In the X-line frame the Hall field has components $B_H(x,z)$ in the y direction which when rotated by the sweep-back angle has components $(B_{Hr}, B_{H\theta}, B_{H\varphi}) = (-B_H \sin\alpha, 0, B_H \cos\alpha)$. Hence, adding the fields due to azimuthal and radial currents we find, $\mathbf{B}(B_r, B_\theta, B_\varphi) = \left\{ B_{r0} \tanh\frac{-z}{D} - B_H \sin\alpha, B_{\theta 0}, B_{\varphi 0} \tanh\frac{-z}{D} + B_H \cos\alpha \right\}$ where we have simply modelled the radial and azimuthal currents with Harris current sheets. Applying this to our transformation (eq. 1) we obtain:

$$\begin{pmatrix} B_x \\ B_y \\ B_z \end{pmatrix} = \begin{pmatrix} \cos\alpha & 0 & \sin\alpha \\ -\sin\alpha & 0 & \cos\alpha \\ 0 & -1 & 0 \end{pmatrix} \begin{pmatrix} B_{r0} \tanh\frac{-z}{D} - B_H \sin\alpha \\ B_{\theta 0} \\ B_{\varphi 0} \tanh\frac{-z}{D} + B_H \cos\alpha \end{pmatrix} \qquad 2$$

$$\begin{pmatrix} B_x \\ B_y \\ B_z \end{pmatrix} = \begin{pmatrix} B_{r0} \tanh\frac{-z}{D} \cos\alpha - B_H \sin\alpha \cos\alpha + B_{\varphi 0} \tanh\frac{-z}{D} \sin\alpha + B_H \sin\alpha \cos\alpha \\ -B_{r0} \tanh\frac{-z}{D} \sin\alpha + B_H \sin\alpha \sin\alpha + B_{\varphi 0} \tanh\frac{-z}{D} \cos\alpha + B_H \cos\alpha \cos\alpha \\ -B_{\theta 0} \end{pmatrix} \qquad 3$$

which simplifies to:

$$\begin{pmatrix} B_x \\ B_y \\ B_z \end{pmatrix} = \begin{pmatrix} B_{r0} \tanh\frac{-z}{D} \cos\alpha + B_{\varphi 0} \tanh\frac{-z}{D} \sin\alpha \\ -B_{r0} \tanh\frac{-z}{D} \sin\alpha + B_{\varphi 0} \tanh\frac{-z}{D} \cos\alpha + B_H \\ -B_{\theta 0} \end{pmatrix} \qquad 4$$

Finally, we note that $\alpha=\tan^{-1}(B_\varphi/B_r)$ and hence $B_{r0} \tanh(-z/D) \sin(\alpha) = B_{\varphi 0} \tanh(-z/D) \cos(\alpha)$ so

$$\begin{pmatrix} B_x \\ B_y \\ B_z \end{pmatrix} = \begin{pmatrix} B_{r0} \tanh\frac{-z}{D} \cos\alpha + B_{\varphi 0} \tanh\frac{-z}{D} \sin\alpha \\ B_H \\ -B_{\theta 0} \end{pmatrix} \qquad 5$$

Hence, the Hall field is obtained from the $B_y$ component of the X-line coordinate system.

The sweep-back angle was measured from the magnetometer data between 08 Oct 2006 0000 UT and 0100 UT and found to be equal to -25.87°±4.87° and so a value of -26° was adopted in this study.

## Electron pitch angle distributions near the X-line

Figure S5 shows reconstructed pitch angle distributions (PAD) in each quadrant of the X-line. CAPS/ELS has an instantaneous FOV of 160°×5.2° which is increased to ~160°×200° by a mechanical scanning platform. Each PAD is produced by combining fluxes measured over a single mechanical ~3 minute scan (actuation). Within this period ELS captures spectra at a cadence between 2 and 32 s but for this study the maximum sampling time was restricted to 8s to avoid undetectable aliasing of the PAD. These fluxes were background-subtracted and sorted into 10° wide pitch angle bins and shifted by the (positive) spacecraft potential to remove trapped spacecraft photoelectrons. The raw spectrograms and reconstructed PADs were checked for evidence of aliasing.

In general the PAD is incomplete due to the limited field of view of the instrument. However, four typical PADs were identified in each quadrant of the X-line. Electron PADs in ion diffusion regions in Earth's magnetotail were found to consist of cool ~100 eV electrons flowing towards the X-line carrying the Hall current, and hotter >1 keV electrons flowing away from the X-line associated with acceleration near the X-line[12]. In Figure S5 we can see that due to the restricted field of view, the orientation of the spacecraft, and changes in orientation of the magnetic field, only electrons flowing out of the X-line are visible on the tailward side of the X-line, and electrons flowing towards the X-line are visible on the planetward side of the X-line. The samples in figure S5 (moving anti-clockwise around the figure) were captured at 0117 UT (above the current sheet and planetward of the X-line), 0242 UT (above the current sheet and tailward), 0341 UT (below and tailward), 0431 UT (below and planetward). We can see that the electrons flowing into the X-line are relatively cool with a peak energy near ~400 eV. The electrons flowing out of the X-line are hot about ~2 keV above the current sheet earlier in the interval at 0242 UT, and ~>10 keV below the current sheet later at 0341 UT. These are entirely consistent with hot electrons flowing out of the X-line and cooler electrons flow in towards the X-line and carrying the Hall current, similar to terrestrial observations[12].

## Ion flows before and during the ion diffusion region encounter

Ion flows throughout the interval are difficult to analyse due to a combination of spacecraft rolls, limited viewing, low signal to noise and aliasing of the distributions. Figure S6 shows a time-energy spectrogram of ion fluxes measured by CAPS/IMS, with the electron fluxes and magnetic field for reference. In this figure ion fluxes have been summed over a 32s internal duty cycle of the instrument (an A-cycle) to improve the signal-to-noise and visibility of ion beams as the instrument actuates across the sky – thus relatively narrow ion beams appear as sharp gradients in the time-energy spectrogram. The "pulsing" in the background is correlated with the actuating motion of CAPS and is thought to be produced by a combination of CAPS actuating through a spatially asymmetrical background produced by radiation from Cassini's radioisotope thermoelectric generators, and changes in the shielding of CAPS from this radiation as it actuates relative to the spacecraft platform and other instruments.

In figure S6 the ion fluxes for particular time intervals are presented as a function of look direction around the spacecraft in order to identify the flow direction of the ions. They also enable us to identify what directions about the spacecraft are not visible to the CAPS detector. These are presented in OAS coordinates in a polar projection. The OAS coordinate system is a spacecraft-centred frame where **S** is a vector from the spacecraft to the planet, **O** is a vector which is obtained from **S**×(**Ω**×**S**) and **A** is a vector along **S**×**O** and completes the right-handed set. The panels in figure S7 are presented in polar coordinates where the polar angle $\theta_{OAS}$ is the angle between a look vector and **S** such that $\theta_{OAS}$=0° represents a direction towards Saturn from Cassini, whereas 90° is perpendicular to the Cassini-Saturn line. The azimuthal angle $\varphi_{OAS}$ is an angle around the **S**. Thus, each panel in figure S7 is drawn from the perspective of an observer on the spacecraft. The centre of the panel is looking at Saturn ($\theta_{OAS}$=0°), the inner circle is $\theta_{OAS}$=90° and the outer circle $\theta_{OAS}$=180°. Hence, ion fluxes in the inner circle are coming from "in front" of the spacecraft, and between the outer and inner circles come from "behind" the spacecraft. Fluxes from the left-hand side of the panel have a component of the flow in a prograde (corotational) direction, and from the right-hand side have a component of the flow in an anti-corotational direction. Fluxes in the upper (lower) half of the panel are coming from above (below) and thus have a flow component directed downwards (upwards). The orange circle indicates the direction of the Sun and the green square shows the direction of corotation.

The ion fluxes in S6 show significant fluxes from 2000 UT on 07 October to 0020 UT on 08 October with a decrease in flux from 2245 to 2330 UT which is correlated with a drop in the electron flux and an increase in the magnitudes of the $B_r$ and $B_\varphi$ components of the magnetic field and the magnetic field strength. Throughout this period the field of view of IMS covers close to the corotation direction and so this drop in flux is consistent with the motion of the spacecraft into the near-lobe – although a rotation of the flow to more azimuthal direction and/or narrowing of the ion beam (faster flows and/or colder ions) cannot be ruled out. Figure S7a shows the ion flow directions from 20:08:18 to 20:11:45 on 07 October and although CAPS does not fully capture the corotation direction, the flows are generally corotational. The ion distributions show clear evidence of two energy peaks, centred on ~300 eV/e and ~4000 eV/q, associated with $H^+$ and $W^+$ respectively, where the ratio in counts $W^+/H^+$= 0.72±0.06 from a fit to CAPS/IMS time-of-flight data.

From 0242 to 0251 UT energetic ion fluxes are observed, coincident with Cassini entering the northern part of the plasma sheet from the near lobe-regions. Figures S7b-S7d show the directions of these fluxes. Although the fluxes are very weak, close to the signal-to-noise threshold of IMS, the flow direction can be determined. At 0242-0245 (S7b) the ions are flowing in a tailward and slightly anti-corotational direction, then appear to be flowing tailward and slightly northward (S7c/S7d). The weakening in the fluxes in S7d is caused by the ions increasing to higher energies (as can be seen in figure S6). Generally, the typical ion energy is above ~2 keV/q and extends to the upper energy/charge range of the instrument. From the time-energy spectra there is some evidence in the beam in S7c for two ion peaks, one at ~8 keV/q and another at ~20 keV/q. From an analysis of the time-of-flight data, the 8 keV/q beam is associated with $H^+$ and the 20 keV/q beam with a species with mass/charge 2 (either $He^{++}$ or $H_2^+$). An 8 keV/q $H^+$ ion has a flow speed of 1200 km s$^{-1}$. This is probably an upper limit to the speed due to the peak energy being due to a combination of bulk and thermal kinetic energy. The ratio of the mass/charge=2 counts to $H^+$ is 7±1. There are no $W^+$ ions to within the error of the analysis, although a $W^+$ ion moving at 1200 km s$^{-1}$ has an energy/charge of 130 keV/q, well above the range of the CAPS/IMS sensor. The energetic ion detectors on Cassini are not orientated in a favourable direction to observe these ions at this time.

The energy spectrum associated with S7d is found about 10 keV/q, which corresponds to a speed of <=1400 km s$^{-1}$. Over the period in the region tailward of the X-line (0146-0355) the CAPS FOV is close to corotation (within ~10-20º) but no measurable fluxes are found in that direction.

From 0354 to 0825 UT the spacecraft undergoes continuous rolling, with another small roll from 0940 to 1000 UT. Due to this rolling behaviour IMS scans rapidly across the sky and it is very difficult to determine the flow directions of the ions. Very narrow features are found in anodes 6/7 at 0401 UT and anodes 1/2 at 0410 UT but these are not visible in OAS plots. This of large-scale flow features is consistent with the planetward-looking FOV and expected planetward reconnection exhaust jets. Evidence for corotational, but slightly tailwards flow is found from 0445 UT onwards, but only sporadic samples (S7e and S7f) are available due to the spacecraft roll. After 0500 UT the spacecraft samples the corotation direction very infrequently, but very low ion fluxes are expected due to the low plasma density, as indicated by the electron measurements[13].

Hence, these observations show that in the tailward region of the diffusion region (as determined from the magnetometer data) CAPS observes a <~1200 km s$^{-1}$ ion beam flowing tailward as expected). By plotting the peaks in ion flux with the look direction information we could determine the flow directions in KSM coordinates and we find the following unit vectors for the three ion beams in figures S7b-S7d: (-0.95, -0.18, -0.27), (-0.77, 0.57, -0.30), and (-0.96, -0.077, -0.28) hence showing an ion beam moving tailward.

## Flux ropes and secondary islands

Plasmoids, with a loop-like or fluxrope structure, are a common signature in planetary magnetotails[14,15] and can be found either travelling planetward or tailward. They can also be seen adjacent to an X-line as a "secondary island" produced as a result of instabilities that set in once reconnection has commenced[16], usually in the presence of a significant guide field (perpendicular to the plane of the X-line). The signature of a plasmoid passing over the spacecraft is a bipolar feature in $B_z$ with deflection in the $B_x$ component in the X-line coordinate system. If the plasmoid has an axial field then it is often termed a flux rope and $B_y$ will show a maximum closest to the centre of the flux rope axis.

By searching for these perturbations in $B_z$ and $B_x$ O-lines have been found in the tailward region of the X-line. Figure S8 shows two periods in the tailward region of the X-line where O-lines have been found. The heavy vertical lines show the passage of the O-line – but not necessarily through the exact centre of the structure. No evidence for flux rope-type signatures are found in these data. The presence of plasmoids close to the X-line is indicative of secondary islands.

## Reconnection rate and Hall field strength

The ratio of the Hall field ($B_y$) to the field in the current sheet ($B_x$) is a dimensionless estimate of the strength of the Hall field. Estimates of the dimensionless strength of the Hall field at Mars show peak values ranging between 0.29 and 0.76 but typically ~0.5 (15). These amplitudes were found to be comparable in size to the dimensionless amplitude of the Hall field at Earth with average values of 0.39±0.16 (45).

Similarly the ratio between the normal field ($B_z$) and the main field ($B_x$) is an estimate of the reconnection rate. For Mars, values ranging between 0.072 and 0.335 with an average of 0.16 and standard deviation of 0.09 have been reported, indicating that reconnection was in the regime of fast reconnection[17]. These values were slightly higher than at Earth but were perhaps the result of a bias towards intense events in the Mars data set.

Figure S9 shows estimates of the strength of the Hall field, $|B_y|/\max(|B_x|)$, and the reconnection rate $|B_z|/\max(|B_x|)$ for the diffusion region encounter described in this paper. The mean value of the Hall field (figure S9e) was 0.18±0.15, although the peak of 0.83 is much higher, compatible with the upper end of the published range[17,18]. The reconnection rate (figure S9f) was found to be 0.13±0.10 with a peak of 0.66 – hence demonstrating fast reconnection – and is similar to martian and terrestrial values.

## Reconnection restart

Sporadically from 0605 UT and onward from 0640 UT on 08 October there is evidence that reconnection restarts or that a fresh part of the plasma sheet moves over the spacecraft and another X-line forms. The spacecraft is located in the southern extreme of the current sheet (steady $B_r<0$) and apparently on closed field lines (typically $B_\theta>0$). The plasma sheet electrons are hotter than typical[19], with energies between 300 eV and 1 keV.

Around 0610 UT a tailward moving plasmoid is observed from a positive-negative bipolar signature in $B_\theta$ (figure 3) suggesting a reconnection X-line has formed planetward of the spacecraft. From 0640 to 0700 UT hot electrons are observed with an energy of ~ 1 – 10 keV. At 0710 UT a dipolarisation front passes the spacecraft as noted by the peak in |**B**| and appearance of hot >1 keV electrons. Another front passes the spacecraft at ~0810 UT. These dipolarisation front passages are interspersed with intervals in the plasma sheet suggesting that a section of the plasma sheet tailward of the spacecraft is reconnecting and Cassini is sporadically immersed in the exhaust from that X-line. After the dipolarisation front at 0810 UT the spacecraft is immersed in hot electrons that increase in energy with time. Additional smaller-scale positive-negative bipolar $B_\theta$ structures are seen in this hot exhaust region suggesting the presence of multiple small-scale dipolarisation fronts[20].

Figure S10 shows ion and electron time-energy spectrograms during these dipolarisation fronts. As noted in section 6, from 0354 to 0825 UT the spacecraft is continuously rolling, with another small roll from 0940 to 1000 UT. Due to this rolling behaviour IMS scans rapidly across the sky and it is very difficult to determine the flow directions of the ions. No significant ion fluxes are observed during the passage of the tailward plasmoid at 0640 UT even though the IMS field-of-view is sufficient to observe tailward flows. During the dipolarisation front at 0710 UT the field-of-view could have seen inward flows from the dawn sector but not from the near-corotation direction.

Significant fluxes are observed between ~0730 and ~0800 UT. Figure S11 shows ion fluxes organised in OAS coordinates. Figures S11a and S11b show ion fluxes from the end of the energetic electron interval after the first dipolarisation front and the entry into the plasma sheet region around ~0730 UT. Figure S11a shows ion fluxes whilst still in the energetic electron region. The IMS field-of-view does not fully capture these ions but assuming IMS captures the edge of the ion beam they appear to be moving inwards and from the duskward direction. Figure S11b shows the next slice and where flows appear from the corotation direction. The nominal plasma sheet during this region has ratios of total counts of various species, $W^+/H^+=13\pm3$ and $(m/q=2)/H^+=12\pm1$, showing a plasma sheet dominated by heavy ions.

No significant ion fluxes are observed between 0800 UT and ~1100 UT, but the IMS viewing is biased to seeing outflows, hence this is not unexpected since the spacecraft is embedded in heated plasma on closed field lines and so we might expect inflows. Figure S12 shows the ion and electron fluxes for the remainder of the dynamical effects on 08 October. Ion fluxes as a function of the field-of-view are in figures S11c-S11i. At 1115 UT (figure S12c) ions >~5 keV/q (speed ~1000 km/s for $H^+$ ions) are observed moving northward, planetward and dawnward consistent with a location in this energised region on closed field lines connected to the exhaust from a reconnection site. Shortly after that (at ~1130) the spacecraft enters the plasma sheet with ~200 eV electrons and ions flowing in the corotation direction (and slightly upward) (figure S12d). The spacecraft re-enters the hot exhaust region around 1240 UT and no significant ion fluxes are seen until 1332 UT despite IMS seeing the whole sky due to spacecraft rolls – although the non-detection of ions might be a combination of flow energies exceeding the energy range of IMS and the flux of ions being below the sensitivity threshold for IMS[13]. At 1332 UT ions are seen just at the edge of the field of view of IMS and suggest inward flow possibly with a downward and dawnward component (figure S11e), again consistent with the location of the spacecraft in the hot exhaust region. Shortly after at 1336 UT the ion flows are more corotational but still with an inward component (figure S11f). Between ~1530 and 1730 the spacecraft is located in the southern lobe, and energetic electron boundary layers are seen near the boundary between the lobe and the plasmasheet. In the boundary layers, ions are found flowing along the magnetic field with pitch angles of 0º (figures S11g and S11h) towards the planet. These boundary layers are on closed field lines, as indicated by the presence of an energetic electron population flowing towards the planet with a pitch angle of 0º, with a counterstreaming component as far as can be seen in the antiparallel direction (figure S13). Finally, the interval ends with a return to the plasma sheet and corotational ion flow (figure S11i).

## Supplementary Materials References

## Figures and captions

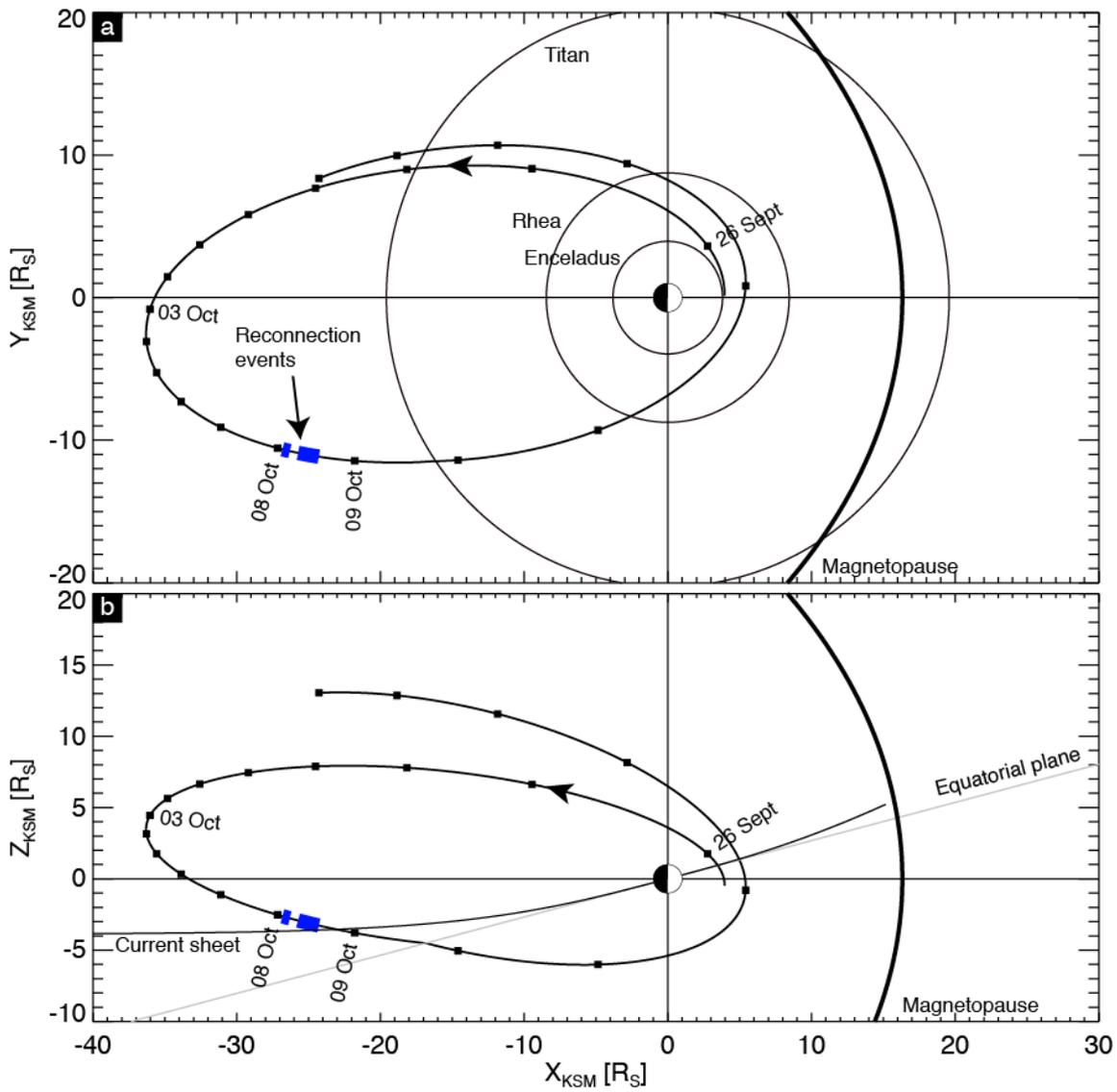

Figure S1: Trajectory of Cassini in KSM during the event in this paper (highlighted in blue). Panel (a) shows the trajectory projected into the X-Y plane and (b) the X-Z plane. The model current sheet location is shown in panel (b) and a model magnetopause in both panels.

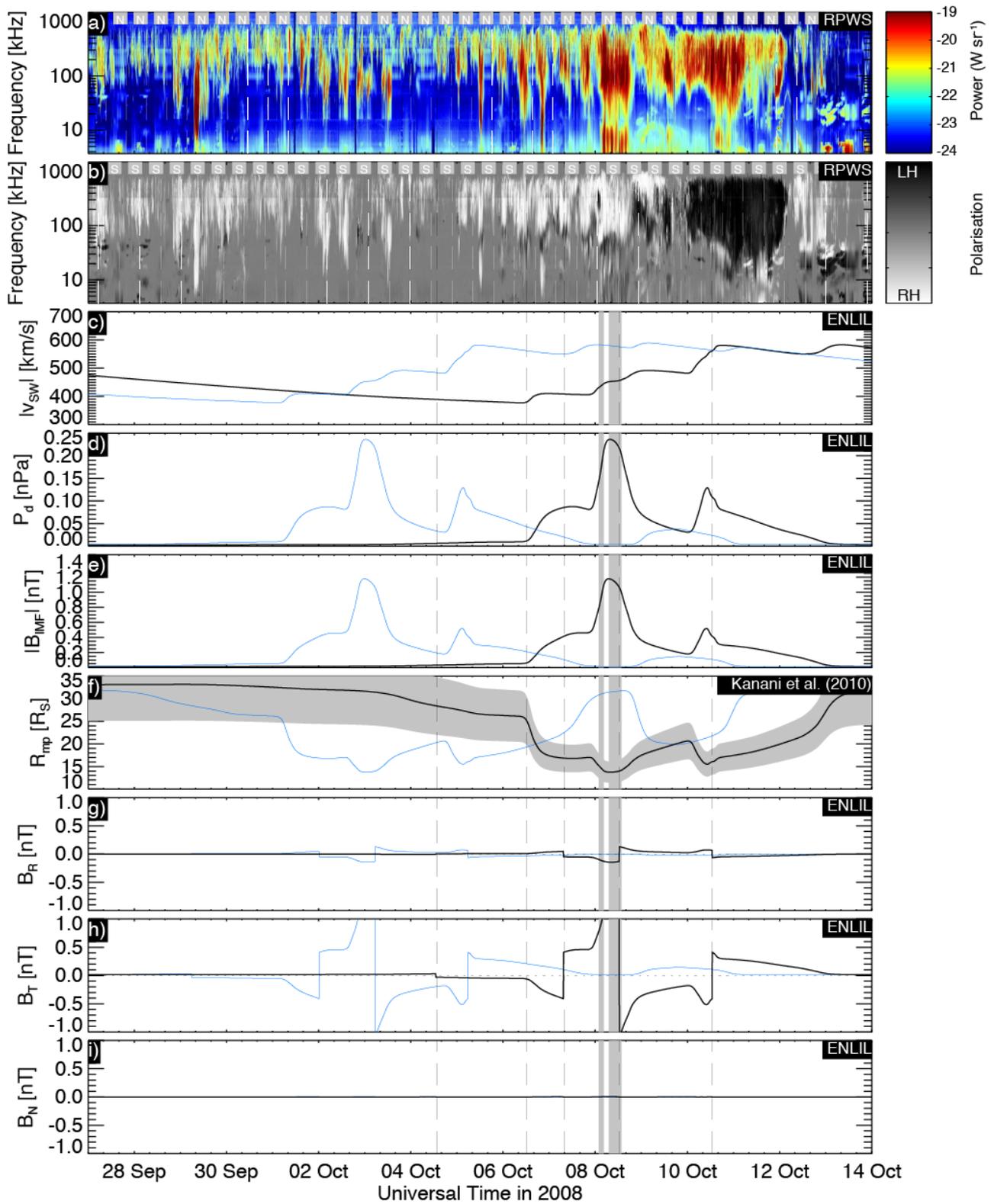

Figure S2: Cassini radio and plasma wave observations and ENLIL solar wind simulation results showing the inferred upstream solar wind conditions during the event: (a) electric field flux density measured by the Cassini/RPWS instrument and scaled to 1 AU distance,

grey "S" and white vertical lines indicates when SKR emissions from the southern hemisphere should be detected based on the SLS4 system; (b) electric field circular polarisation measured by the Cassini/RPWS instrument (white indicates emissions from the northern hemisphere, black from the south), grey and white vertical lines indicates when SKR emissions from the northern hemisphere should be detected based on the SLS4 system; (c) solar wind speed from ENLIL; (d) solar wind dynamic pressure from ENLIL; (e) interplanetary magnetic field strength from ENLIL; (f) inferred subsolar position of the magnetopause based on the ENLIL dynamic pressure and a model magnetopause[9]; (g-i) magnetic field in the RTN coordinate system from ENLIL. The vertical dashed black lines indicate HCS crossings. The grey vertical bars indicate the reconnection regions in Figure 3 of the main manuscript. In each ENLIL panel the blue curves show the original ENLIL data, black shows the ENLIL data which has been shifted in time by 5.3 days to match the enhancements in the measured SKR flux, as discussed in the SOM text.

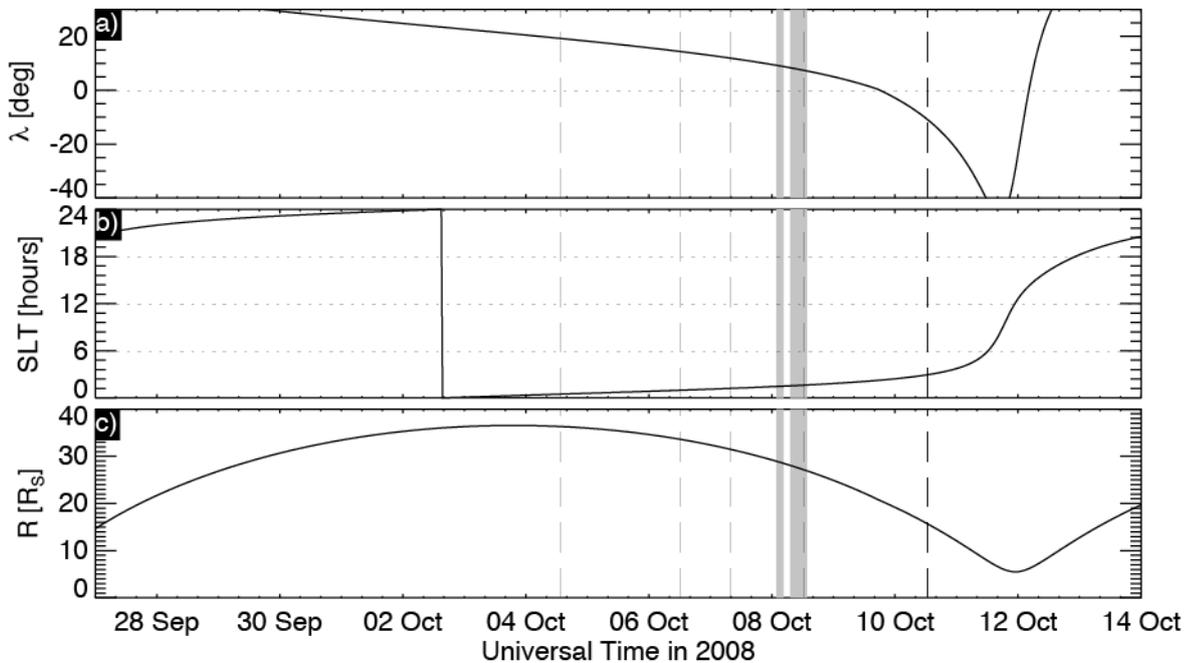

Figure S3: Cassini orbital parameters used to interpret Cassini radio and plasma wave observations: (a) latitude, (b) local time and (c) radial distance of Cassini.

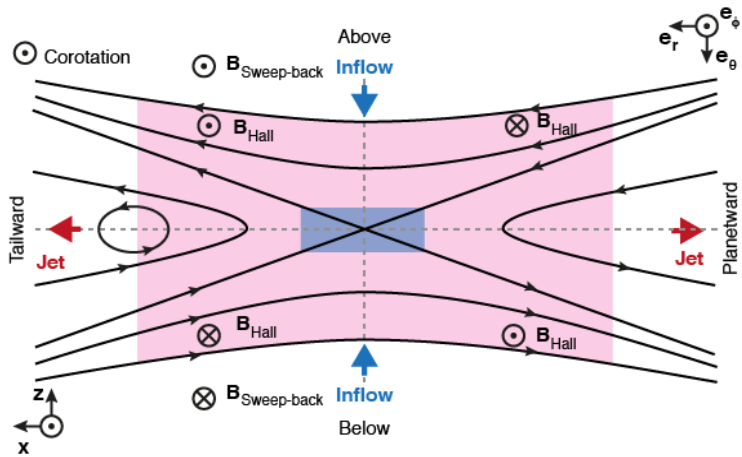

Figure S4: Schematic diagram showing the reconnecting current sheet with the ion (pink) and electron (blue) diffusion regions[11], inflow and outflow jets, and the orientation of the Hall field and magnetic field associated with the sweep-back of the magnetic field.

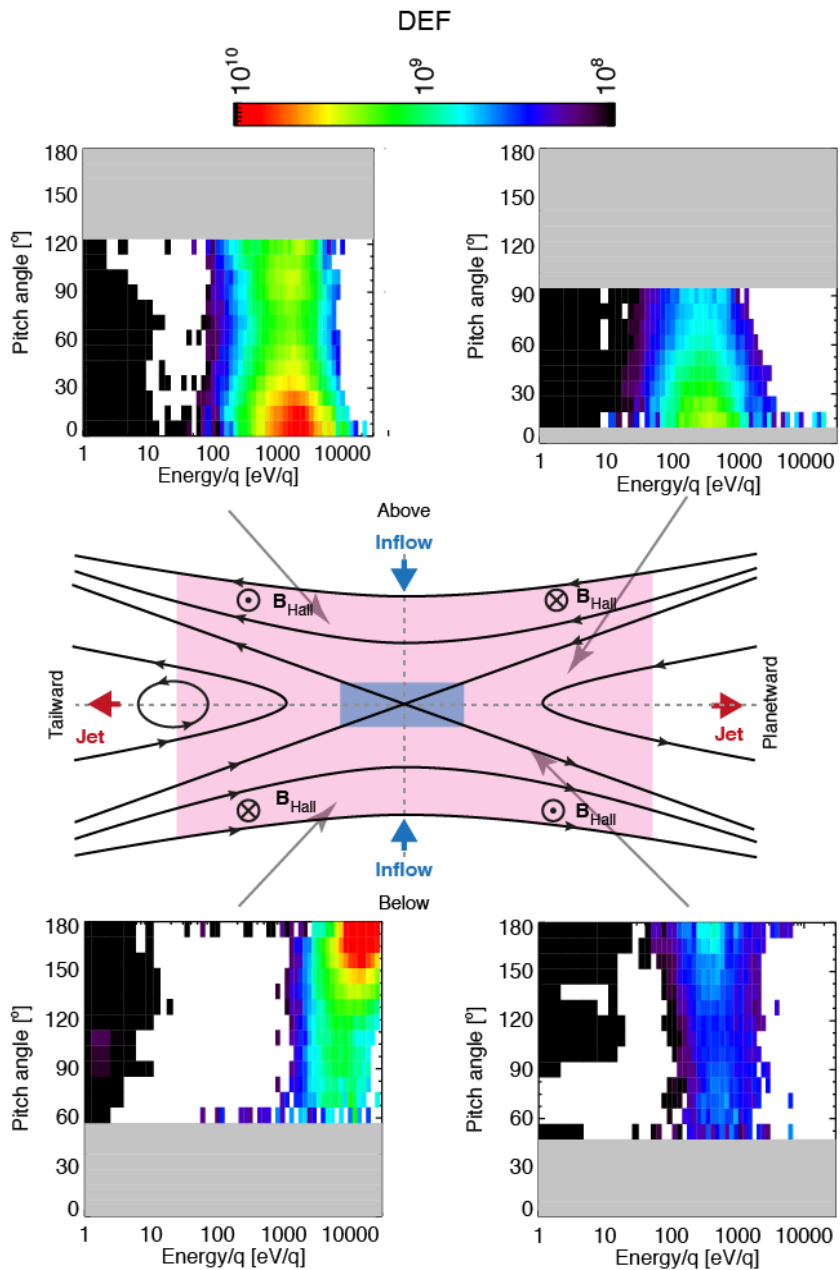

Figure S5: Electron pitch angle distributions in each quadrant of the X-line. Because of changes in orientation of the spacecraft and the magnetic field, combined with the 160°×5° instantaneous field of view of the ELS analyser, the pitch angle coverage is generally incomplete with pitch angles of only 0° or 180° covered by the instrument field of view. The colour scale shows the measured differential energy flux in units of eV m$^{-2}$ s$^{-1}$ sr$^{-1}$ eV$^{-1}$.

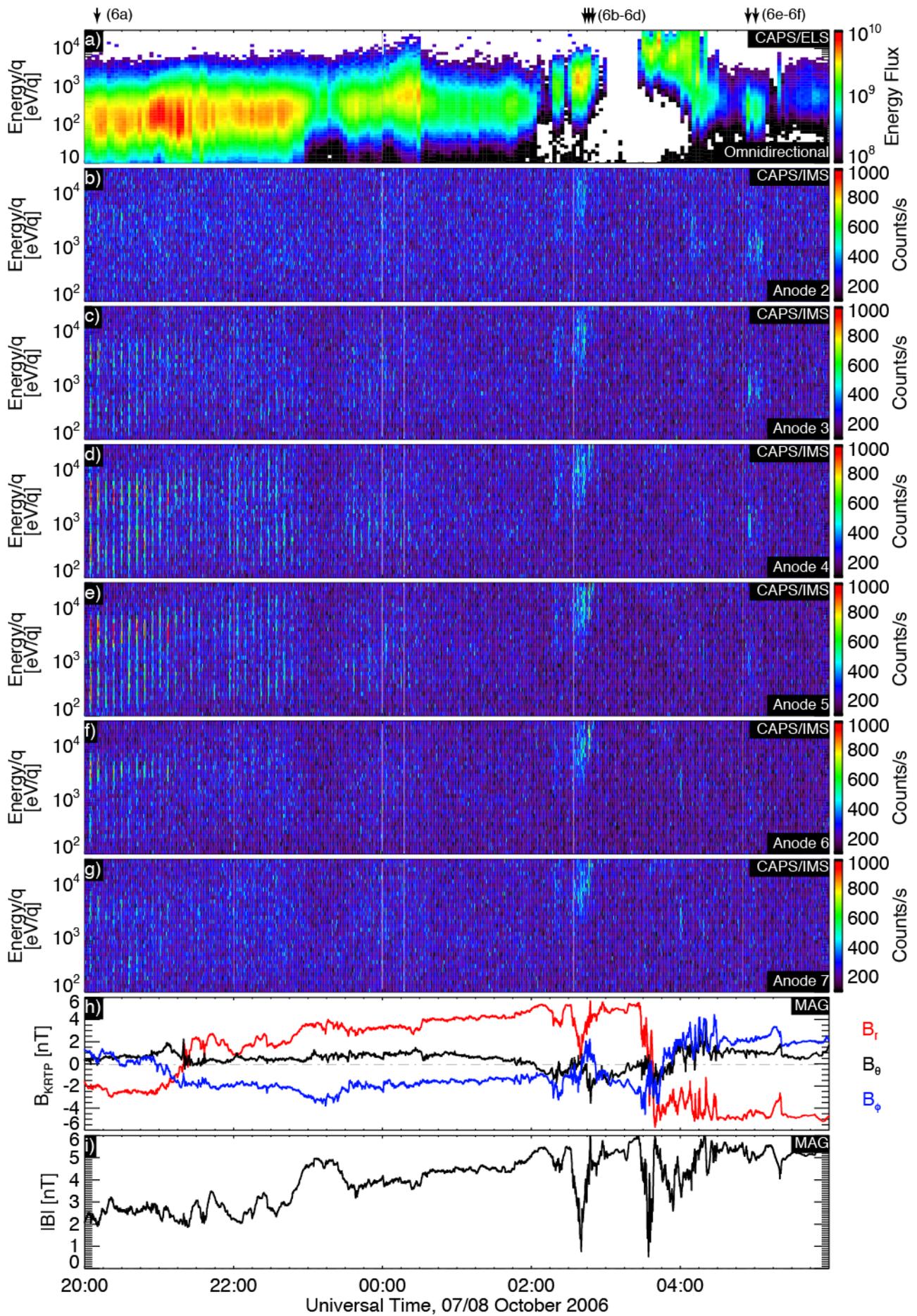

Figure S6: Ion fluxes measured by CAPS/IMS with electron fluxes and magnetic field data for reference. Panels (b-g) show ion fluxes from anodes 2-7 of CAPS/IMS on a linear colour scale from 100 to 1000 counts/32s (summed over a 32s instrument duty cycle). There are no measurable fluxes below 100 eV/q. The arrows at the top of each panel indicate the times of the OAS plots presented in figure S8.

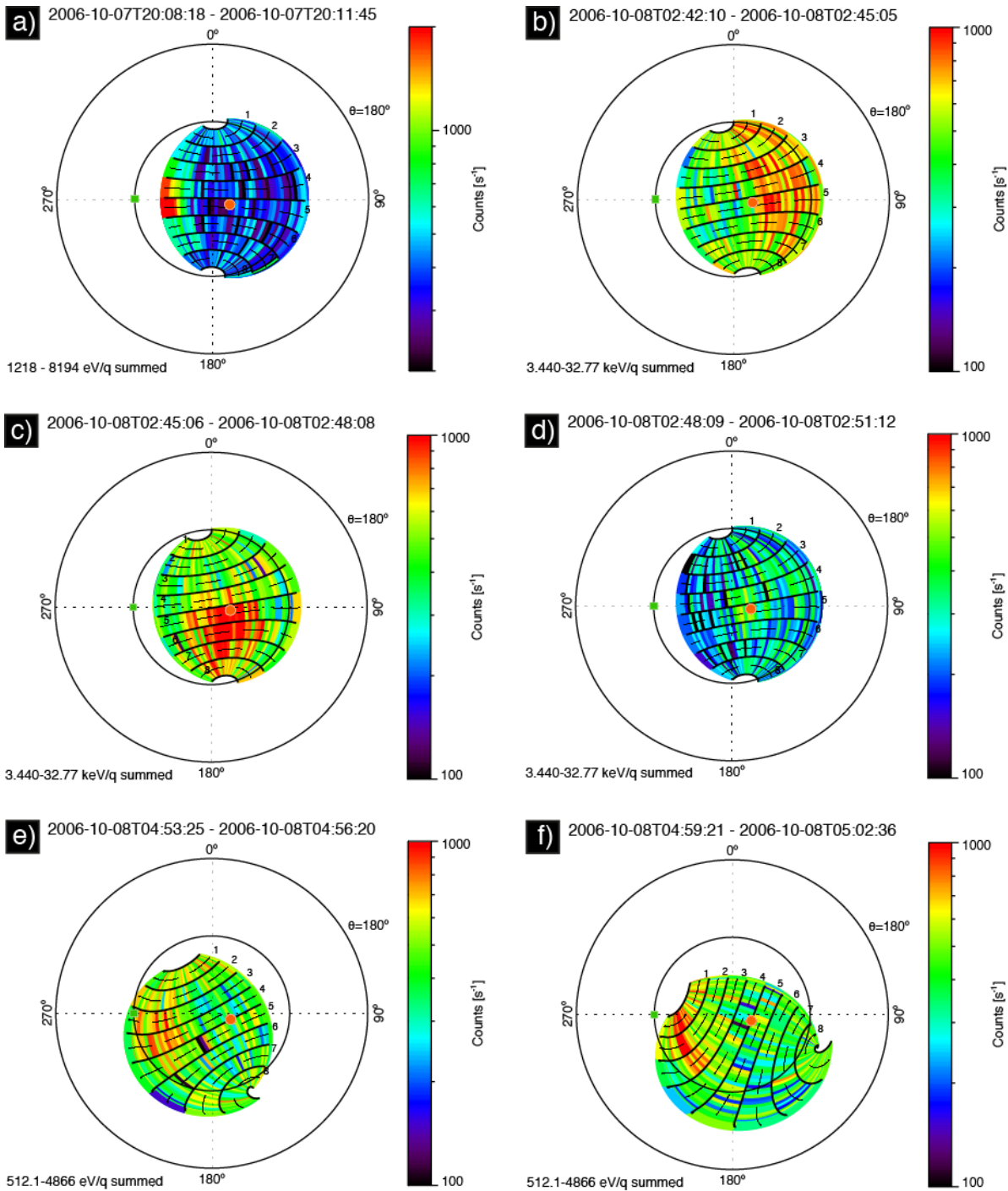

Figure S7: Ion fluxes presented as a function of look direction in OAS coordinates. Red and blue symbols show 0º and 180º pitch angle directions, orange circles show the Sun direction, and green square shows the corotation direction. Saturn is in the centre of each panel.

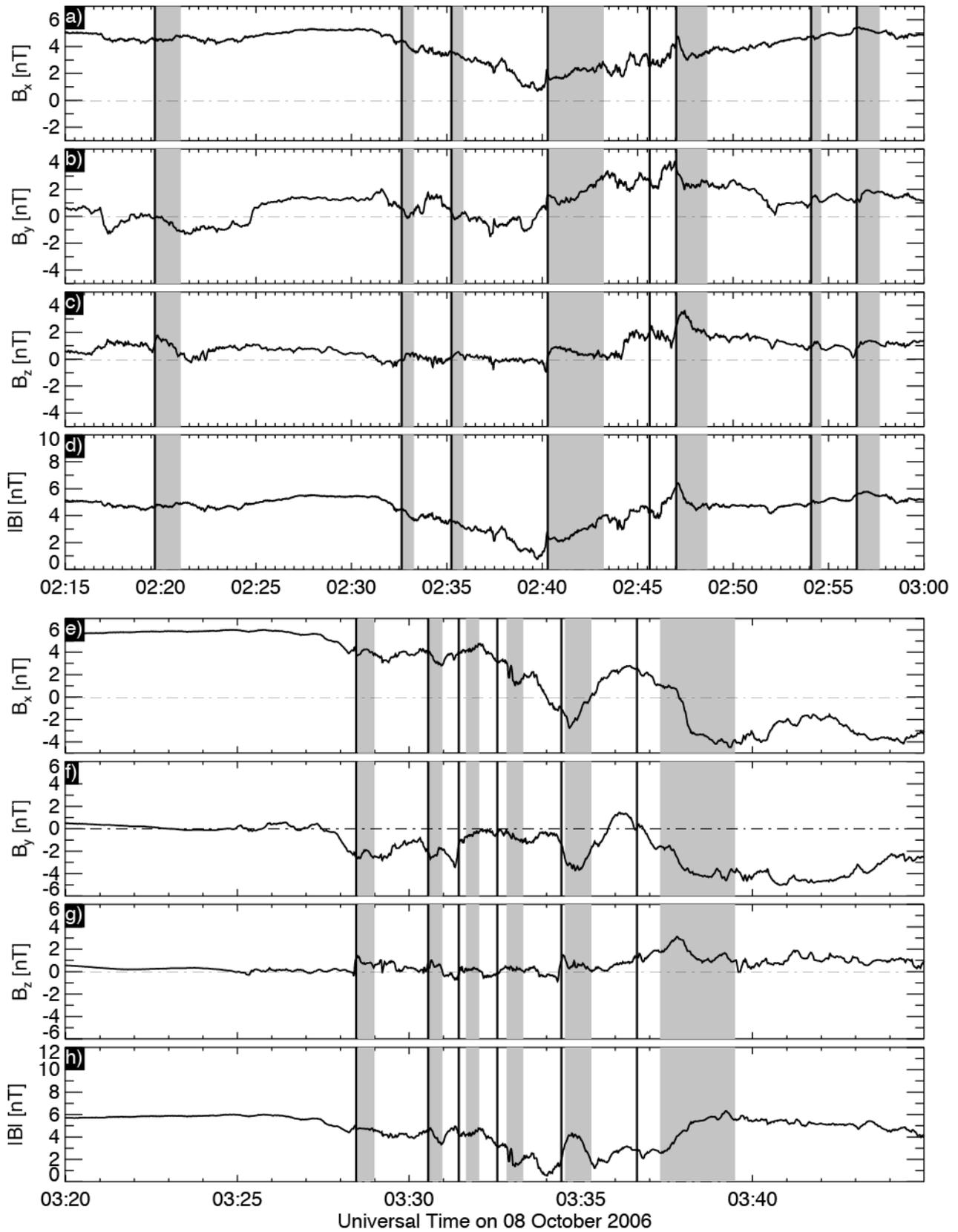

Figure S8: Small plasmoids observed in magnetometer data on the tailward side of the X-line. Panels (a-d) show magnetometer data from 0215 – 0300 UT and panels (e-h) show

data from 0320-0345 UT on 08 October. Both sets of data are presented in the X-line coordinate system. The bold vertical lines indicate the passage of small plasmoids, the shaded grey regions indicate post-plasmoid plasma sheets. Note the different time scales and y-axis scales in each plot

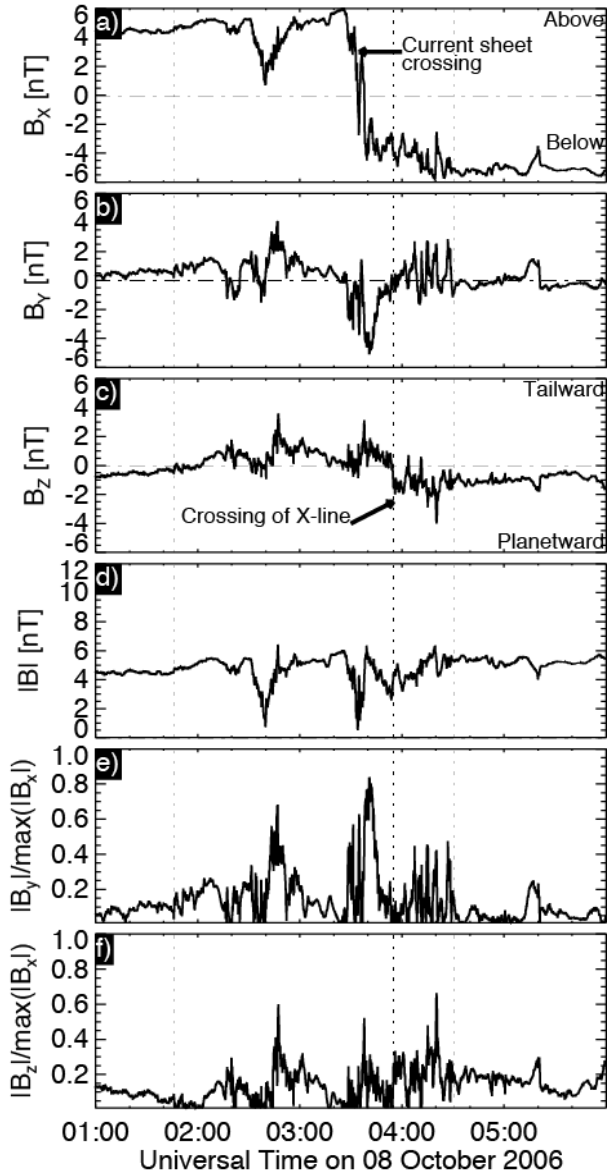

Figure S9: Reconnection rate and Hall field strength estimates near the diffusion region. Panels (a-d) show the measured magnetic field in the X-line coordinate system, panel (e)

shows the strength of the Hall field expressed as the dimensionless ratio $|B_y|/\max(|B_x|)$ and panel (f) shows a dimensionless proxy for the rate of reconnection given by as the dimensionless ratio $|B_z|/\max(|B_x|)$.

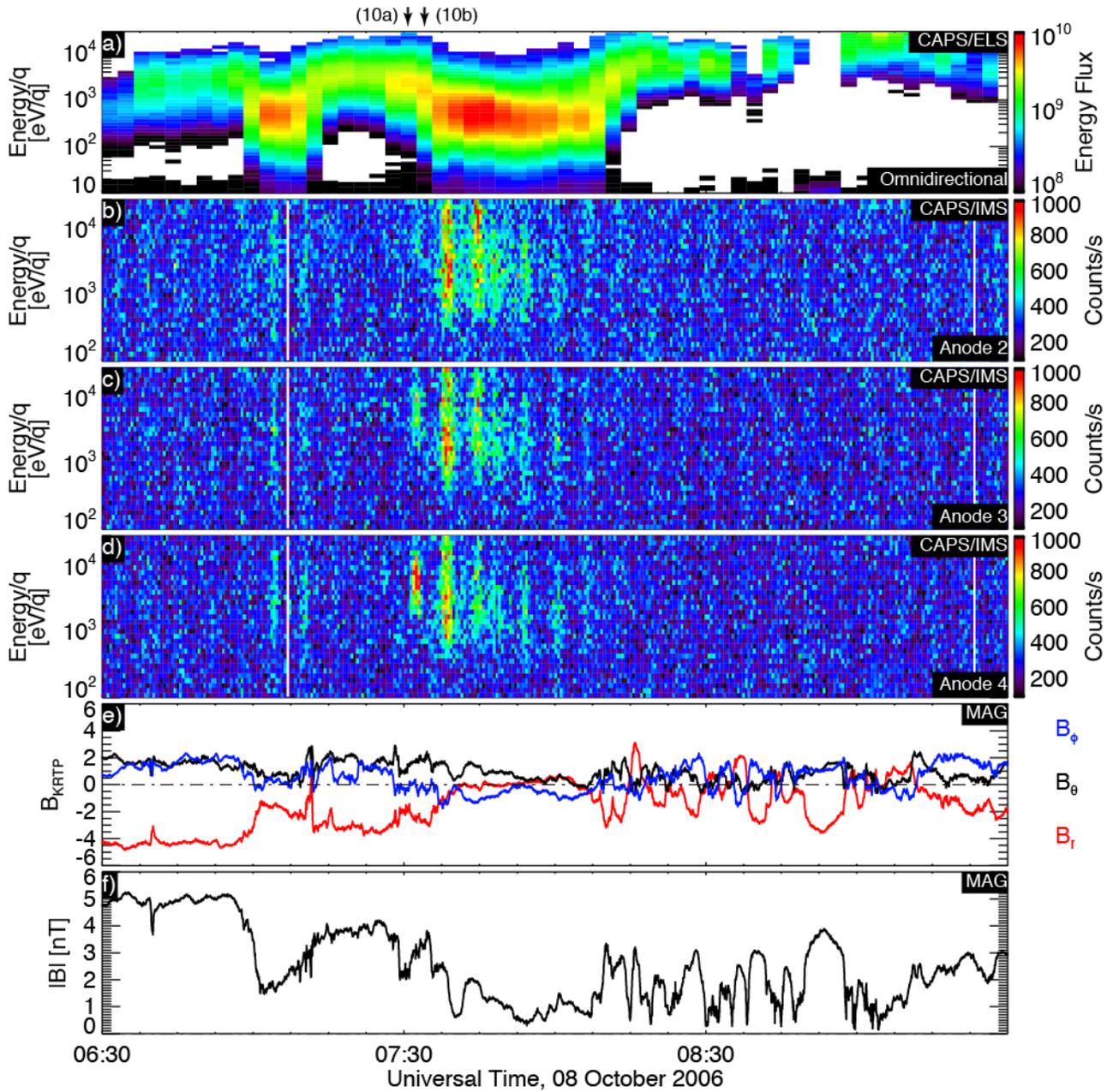

Figure S10: Electron, ion and magnetic field observations during re-encounter or restart of reconnection. Panel (a) shows a CAPS/ELS time-energy spectrogram of omni-directional

flux averaged over a CAPS actuation cycle. Panels (b-d) show time-energy spectrograms of ion flux averaged over 32s from anodes 2-4 of CAPS/IMS (the anodes showing the highest flux). Panels (e) and (f) show the magnetic field components and field magnitude. The arrows at the top of each panel indicate the times of the OAS plots presented in figure S11.

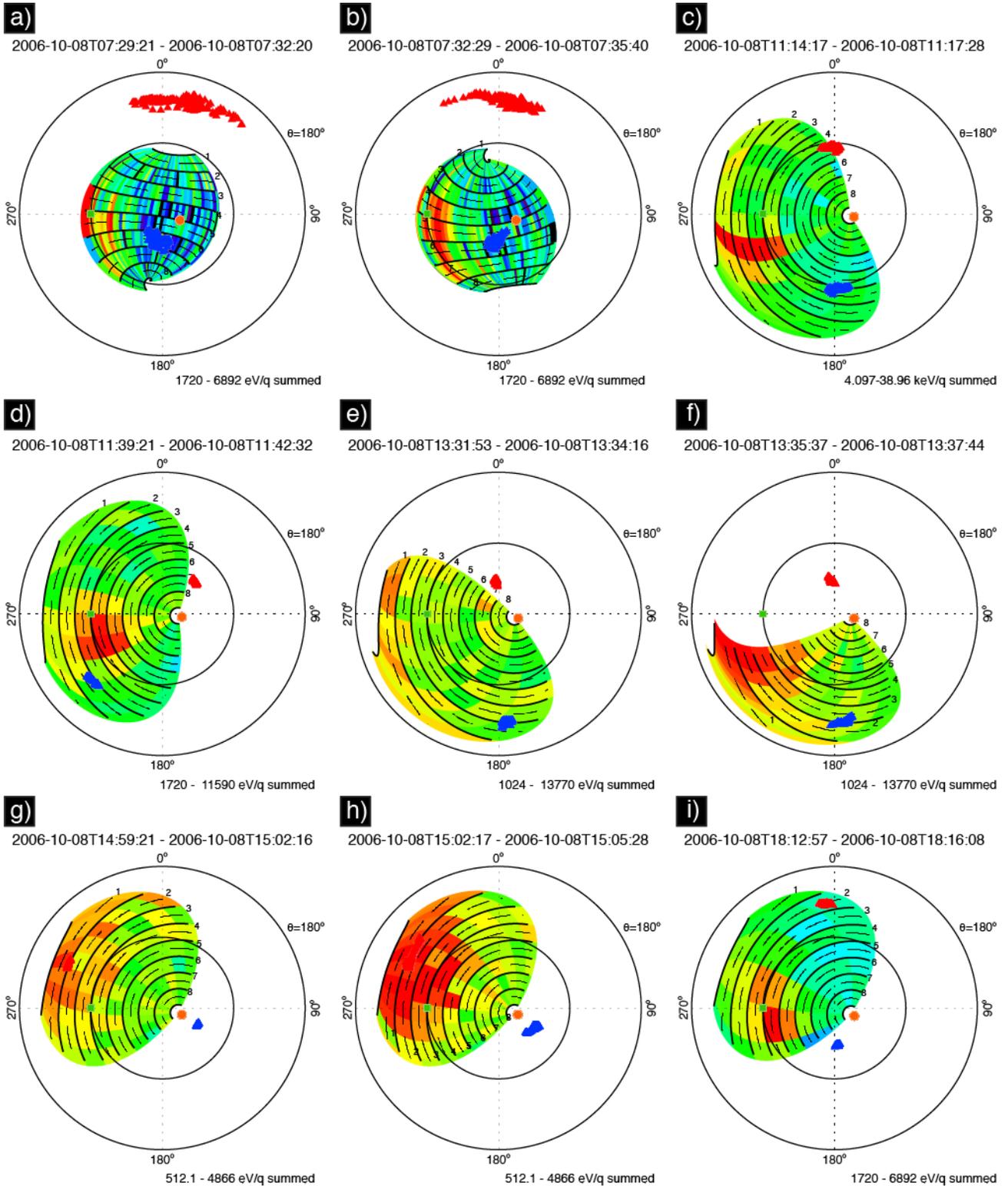

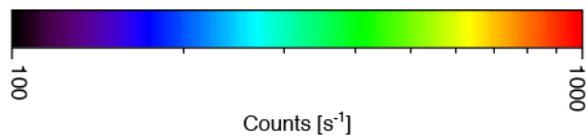

Figure S11: Ion fluxes presented as a function of look direction in OAS coordinates corresponding to times in figures S9 and S11. Red and blue symbols show 0º and 180º pitch angle directions, orange circles show the Sun direction, and green square shows the corotation direction. Saturn is in the centre of each panel.

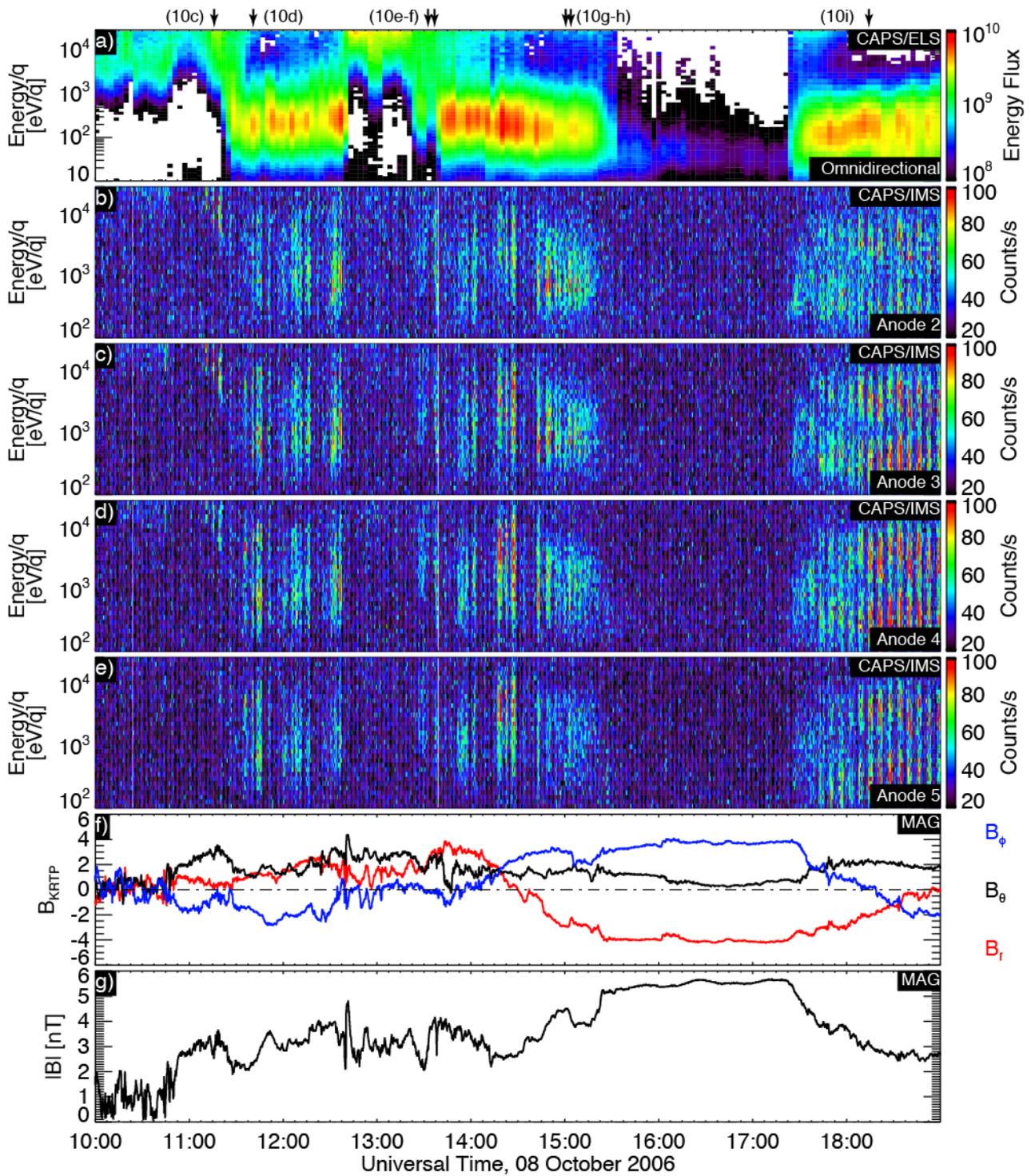

Figure S12: Electron, ion and magnetic field observations during re-encounter or restart of reconnection. Panel (a) shows a CAPS/ELS time-energy spectrogram of omni-directional flux averaged over a CAPS actuation cycle. Panels (b-e) show time-energy spectrograms of ion flux averaged over 32s from anodes 2-5 of CAPS/IMS (the anodes showing the highest flux). Panels (f) and (g) show the magnetic field components and field magnitude. The arrows at the top of each panel indicate the times of the OAS plots presented in figure S11.

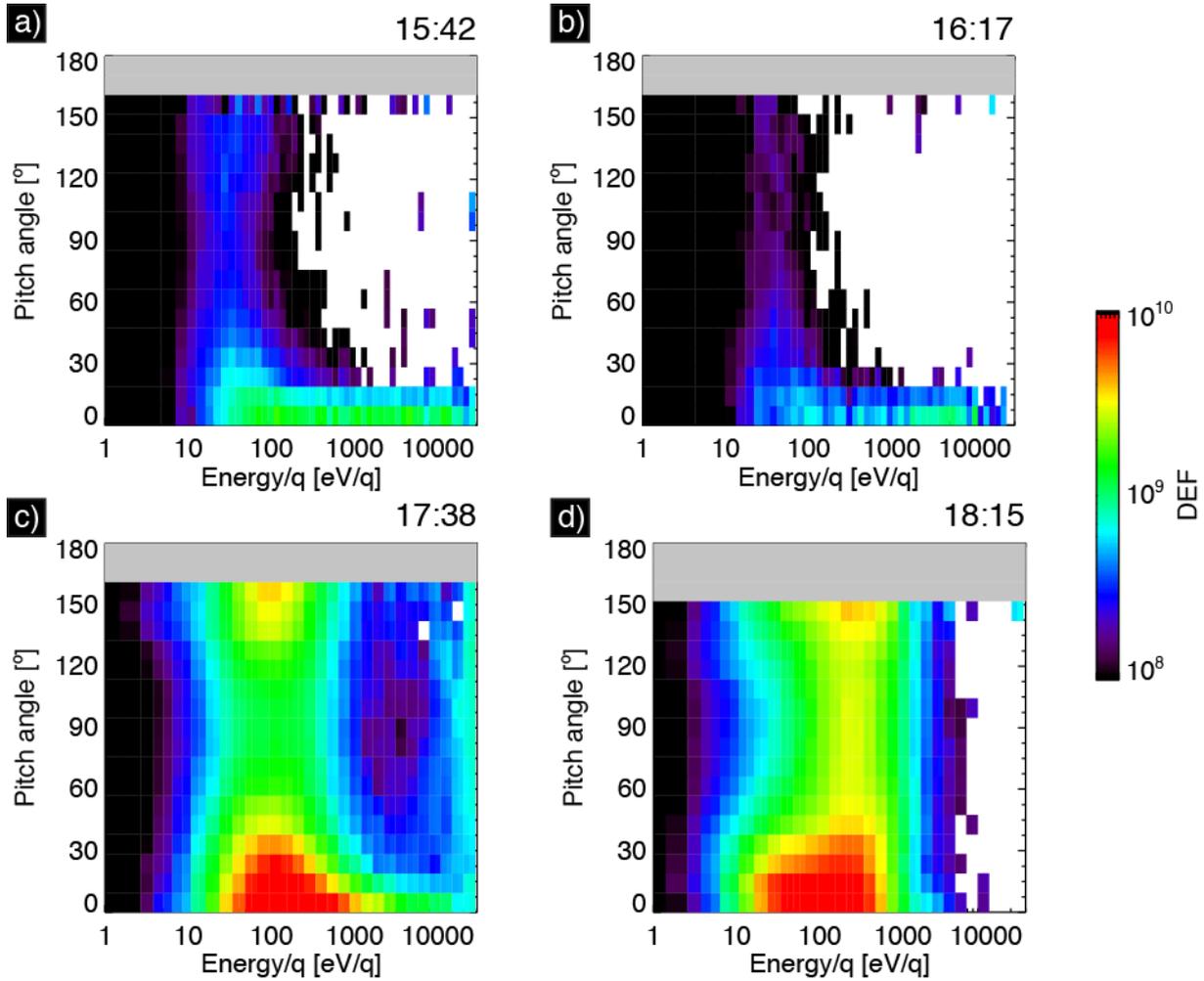

Figure S13: Electron pitch angle distributions near the lobe showing electrons forming a beam flowing parallel to the magnetic field (0º pitch angle) near the lobe/plasma sheet boundary (a and c), in the lobe (b), and returning to a bidirectional ~100 eV population in the plasma sheet.